# Atomic-scale imaging of $CH_3NH_3PbI_3$ structure and its decomposition pathway

Shulin Chen[1,2], Changwei Wu[3], Bo Han[1], Zhetong Liu[1], Zhou Mi[4], Weizhong Hao[4], Jinjin Zhao[4,*], Xiao Wang[3,*], Qing Zhang[5], Kaihui Liu[6,7], Junlei Qi[2], Jian Cao[2], Jicai Feng[2], Dapeng Yu[8], Jiangyu Li[3,9,10,*], Peng Gao[1,7,11,*]

[1]Electron Microscopy Laboratory, International Center for Quantum Materials, School of Physics, Peking University, Beijing 100871, China

[2]State Key Laboratory of Advanced Welding and Joining, Harbin Institute of Technology, Harbin 150001, China

[3]Shenzhen Key Laboratory of Nanobiomechanics, Shenzhen Institute of Advanced Technology, Chinese Academy of Sciences, Shenzhen 518055, Guangdong, China

[4]School of Materials Science and Engineering, Shijiazhuang Tiedao University, Shijiazhuang, 050043, China

[5]Department of Materials Science and Engineering, College of Engineering, Peking University, Beijing 100871, China

[6]State Key Laboratory for Mesoscopic Physics, School of Physics, Peking University, Beijing 100871, China

[7]Collaborative Innovation Center of Quantum Matter, Beijing 100871, China

[8]Department of Physics, South University of Science and Technology, Shenzhen 518055, China.

[9]Department of Materials Science and Engineering, Southern University of Science and Technology, Shenzhen 518055, China

[10]Guangdong Key Provisional Laboratory of Functional Oxide Materials and Devices, Southern University of Science and Technology, Shenzhen 518055, China

[11]Interdisciplinary Institute of Light-Element Quantum Materials and Research Center for Light-Element Advanced Materials, Peking University, Beijing 100871, China

These authors contributed equally: Shulin Chen, Changwei Wu
*Correspondence and requests for materials should be addressed to: p-gao@pku.edu.cn (P. Gao); lijy@sustech.edu.cn (J. Y. Li); jinjinzhao2012@163.com (J.J. Zhao); xiao.wang@siat.ac.cn (X. Wang)

**Abstract**

Understanding the atomic structure and structural instability of organic-inorganic hybrid perovskites is the key to appreciate their remarkable photoelectric properties and failure mechanism. Here, using low-dose imaging technique by direct-detection electron-counting camera in transmission electron microscope, we investigate the atomic structure and decomposition pathway of $CH_3NH_3PbI_3$ ($MAPbI_3$) at the atomic scale. We successfully image the atomic structure of perovskite in real space under ultra-low electron dose condition, and observe a two-step decomposition process, i.e. initial loss of MA followed by the collapse of perovskite structure into $6H-PbI_2$ with their critical threshold dose also determined. Interestingly, an intermediate phase ($MA_{0.5}PbI_3$) with locally ordered vacancies can robustly exist before perovskite collapses, enlightening strategies for prevention and recovery of perovskite structure during degradation. Associated with structure evolution, the bandgap gradually increases from ~1.6 eV to ~2.1 eV, and it is found that both C-N and N-H bonds can be destroyed under irradiation, releasing $NH_3$ and leaving hydrocarbons. These findings enhance our understanding of the photoelectric properties and failure mechanism of $MAPbI_3$, providing potential strategy into material optimization.

**Introduction**

Organic-inorganic hybrid perovskites (OIHPs) have attracted great research interests as promising materials for the next generation photovoltaic energy harvesting[1,2], electro-optic detection[3,4] and all-optical conversion[5,6]. Their remarkable properties are underpinned by hybrid perovskite atomic structures[7], $ABX_3$, with organic species such as $CH_3NH_3$ (MA) and $CH(NH_2)_2$ (FA) occupying A-site and inorganic Pb in B-site surrounded by X-octahedron formed by halogen elements like I or Br. In particular, the corner-sharing $PbI_6$-octahedron is believed to be beneficial for carrier diffusion[8,9], while its distortion under chemical strain[10] makes the band gap tunable, ideal for photovoltaic conversion. Moreover, the organic cation as well as the hydrogen bonding may lead to spontaneous polarization and ferroelectricity[11], which promotes the separation of photoexcited electron-hole pairs, and thus reduce the recombination and improve the carrier lifetimes[12]. These characteristics are responsible for the promising optoelectronic properties including high carrier mobility, long charge diffusion length and superior power conversion efficiency[13]. Nevertheless, the exact atomic structure of OIHPs is still unsettled, with two possible space groups, polar I4cm and nonpolar I4/mcm still hotly debated depending on the orientations of polar molecules such as MA[14]. While many perovskite oxides are polar with strong ferroelectricity, the polarity of OIHPs has yet to be firmly established[15].

The lack of detailed understanding on atomic structure of OIHPs is largely due to the incapability to image OIHPs at the atomic scale[16,17]. It is well known that OIHPs are quite unstable and prone to decomposition under electron beam irradiation[18-20]. While much progress has been made in transmission electron microscopy (TEM) characterization of OIHPs, direct visualization of atomic structure remains to be elusive. Initial TEM studies at low doses are mainly observing the morphology evolutions[18] and structure transitions by reciprocal-space electron diffraction (ED) techniques[19], and many of the earlier studies mislabeled decomposition product $PbI_2$ as $MAPbI_3$[21-23]. With the help of direct-detection electron-counting (DDEC) camera, high-resolution TEM (HRTEM) image of $CH_3NH_3PbBr_3$ has been successfully obtained, which is much more stable than $MAPbI_3$, though the observed off-centered MA with different orientations has not been substantiated[16]. Recently, low-dose scanning transmission electron microscopy (STEM) technique provides atomic-scale insights into crystalline defects of $CH(NH_2)_2PbI_3$ ($FAPbI_3$), though the obtained atomic structures has already been damaged due to the large doses involved (53-221 e Å$^{-2}$)[24]. Cryo-HRTEM has been used to image $MAPbI_3$ at 100 e Å$^{-2}$, yet the corresponding fast Fourier transform (FFT) pattern lacks (002) reflection, reflecting substantial beam damage[25]. Furthermore, Li et al. found that superstructure reflections, a sign of structural transition due to beam damage, have already appeared at a dose as low as 7.6 e Å$^{-2}$ [26], and under Cryo-TEM, rapid amorphization has also been observed[18,27]. Indeed, the damage-free pristine structure of $MAPbI_3$ has not been imaged at the atomic scale, and the corresponding real-space degradation pathway is yet to be established, thus motivating this study.

It is well known that STEM imaging introduces comparably larger dose and damage than low-dose HRTEM, while the contrast of HRTEM is sensitive to imaging condition, making it difficult to identify the specific atomic columns of $MAPbI_3$[28]. To overcome these difficulties, we adopted DDEC camera combined with imaging technique using a negative value of the spherical-aberration coefficient[29], and we have successfully imaged the atomic structure of $MAPbI_3$ in real space at a dose as low as 0.7 e Å$^{-2}$, ensuring minimum beam damage if any. We further observed a two-step degradation pathway at atomic scale, initiated with the loss of MA to form a superstructure $MA_{0.5}PbI_3$ with ordered MA vacancies ($V_{MA}$), followed by the diffusion of I and Pb to form the decomposed 6H-$PbI_2$, with the corresponding critical doses also identified. During the process, both C-N and N-H bonds can be destroyed under irradiation, releasing NH$_3$ and leaving hydrocarbons. The continuous structure transformations result in gradually increased bandgap, which is confirmed by scanning electron microscope cathodoluminescence (SEM-CL) experiments and validated by density functional theory (DFT) calculations. The direct visualization of structure and degradation process at the atomic scale provide valuable sights into understanding the properties and stability of OIHPs. Furthermore, the emergence of superstructure before the collapse of perovskite framework also points toward a strategy for stabilizing the materials during the degradation.

**Identification of damage-free threshold dose**

MAPbI$_3$ nanocrystals with 10-20 nm size and good crystalline (Supplementary Fig. 1) are chosen for low dose imaging. Using DDEC camera, HRTEM images of MAPbI$_3$ can be acquired at low doses as shown in **Fig. 1**. It is noted that sufficient dose (Supplementary Fig. 2) is needed to obtain image with good quality and superstructure diffraction reflections appear due to the generation of intermediated phases when the dose is larger than 2.7 e Å$^{-2}$ (**Fig. 1b**, **g**). Judging from the corresponding FFT patterns, the [001] MAPbI$_3$ with intermediated phases gradually transforms at 13.6 e Å$^{-2}$, and finally decomposes into [-441] or [48-1] 6H-PbI$_2$ (Supplementary Fig. 3) at 272.0 e Å$^{-2}$. Thus the threshold dose for MAPbI$_3$ without forming superstructures is determined to be 2.7 e Å$^{-2}$ while 272.0 e Å$^{-2}$ results in the complete decomposition into PbI$_2$. These doses identified can guide the future TEM characterizations of MAPbI$_3$.

**Atomic-imaging of MAPbI$_3$ structure and the intermediate phase**

We then investigate the atomic-scale structure via imaging technique using a negative value of the spherical-aberration coefficient (Cs), which has enabled the successful observation of both light and heavy element in oxide perovskite[29]. **Fig. 2a** is the HRTEM image acquired at a negative Cs with an overfocus, wherein the brightest 'I' column, second brightest 'II' column and the darkest 'III' column can be observed. By comparing the atomic structure features of MAPbI$_3$ (**Fig. 2b**) and HRTEM simulations (Supplementary Fig. 4), 'I', 'II' and 'III' atomic columns are identified to be Pb-I, I and MA, respectively. With increased dose, the intensity of MA is decreased at 10.5 e Å$^{-2}$ as shown in **Fig. 2c**. The quantitative intensity analysis (**Fig. 2d**) further verifies that MA intensity decreases within the initial 10.5 e Å$^{-2}$, and then remains stable until 28.0 e Å$^{-2}$ followed by gradual increase (Supplementary Fig. 5). The decreased intensity is caused by the formation of $V_{MA}$[30] while the unchanged intensity is likely resulted from a relatively stable intermediate phase, suggesting the reversible loss of MA and preservation of perovskite structure. The subsequently increased intensity is resulted from the diffusion of I and Pb, as discussed in the following. **Fig. 2e** and Supplementary Fig. 6 further show $V_{MA}$ appears at every other 'III' column, as illustrated in **Fig. 2f**. Such cation vacancy ordered structure with superstructure reflections (Supplementary Fig. 7) corresponds to MA$_{0.5}$PbI$_3$, whose stability is verified by molecular dynamic simulation (Supplementary Fig. 8). Accordingly, it is concluded the loss of MA starts even at 1.0 e Å$^{-2}$ and reaches a balanced state between 10.5 and 28.0 e Å$^{-2}$ to form ordered $V_{MA}$, wherein the perovskite structure framework is preserved.

**Evolution of electronic structure and chemical bonding**

The effect of ordered $V_{MA}$ on its electronic structure is further investigated. **Fig. 3a**, **b** show the calculated band structure of MAPbI$_3$ and MA$_{0.5}$PbI$_3$. The band gap of MAPbI$_3$ is 1.56 eV while it is 1.69 eV for MA$_{0.5}$PbI$_3$. The increased band gap is caused by the enhanced hybridization between I-5p and Pb-6p atomic orbitals and the conduct band minimum shifting about 0.1 eV to high energy level, as explained in Supplementary Fig. 9. To confirm this analysis, we also carried out SEM CL experiments. Supplementary Fig. 10 shows the initial CL emission with a single excitonic peak at the photon energy of 1.58 eV. Time-series CL emissions in **Fig. 3c** show that the observed peaks gradually become broader and shift to higher energy (2.05 eV) with the excitonic peak intensity decreasing. Such blue-shift is associated with the electron-beam-induced phase transformations[31], i.e. forming the intermediated phases and decomposed product 6H-$PbI_2$, considering that the calculated bandgaps of $MAPbI_3$, $MA_{0.5}PbI_3$ and 6H-$PbI_2$ are 1.56, 1.69, and 2.15 eV (Supplementary Fig. 11), respectively, in good agreement with experimental observation.

In addition to electronic structure evolutions, it is worth investigating how the chemical bonding within organic components evolves (**Fig. 3d**, **e**) during the degradation. Vibrational electron energy loss spectroscopy operated at 'aloof' mode[32] (inset of **Fig. 3e**), which enables precise control of damage by changing the distance between electron beam and sample[32], is used to obtain the characteristic vibrational modes of $MAPbI_3$ (**Fig. 3d**). We can observe the vibrational signals of $CH_3$-$NH_3^+$ rock at 113 meV, $CH_3$ bend at 177 meV, and $NH_3^+$ stretch at 391 meV[33]. Time-series vibrational spectroscopy shows that the peaks of $CH_3$-$NH_3^+$ rock and $NH_3^+$ stretch gradually disappear with increased time, suggesting the breakage of C-N and N-H bonds. The extracted intensities of C-N and C-H bonds are shown in **Fig. 3e** and the processing details are shown in Supplementary Fig. 12. It is observed the intensities of C-N bonds gradually decrease suggesting C-N bonds are gradually destroyed by the electron beam. In contrast, the intensities of C-H bond increase likely due to the formation of hydrocarbon[34,35] (Supplementary Fig. 12). Based on these analyses, it is believed that under this 'aloof' beam, C-N and N-H bonds can be destroyed, releasing $NH_3$ gas and leaving hydrocarbon products.

**Atomic-scale observation of I and Pb diffusion**

Based on above study, we further investigate the atomic-scale decomposition pathway of MAPbI$_3$. **Fig. 4a** is the structure of perovskite with V$_{MA}$, as illustrated in **Fig. 4e**. With increased doses, it is observed that the intensities of these three atomic columns gradually change (**Fig. 4b-d**). A quantitative analysis of the intensity changes (Supplementary Fig. 13) shows that the intensities of 'I' columns initially increase and then gradually decrease while the intensities of 'II' columns continuously increase and the intensities of 'III' columns initially decrease and then gradually increase. Finally, intensities of all three types atomic columns converge, indicating the formation of PbI$_2$. The initial intensity decrease of MA results from the formation of V$_{MA}$, while the following increased intensity of MA column and decreased intensity of Pb-I column are believed to be caused by two kinds of Pb-I diffusion as illustrated in **Fig. 4f**. One is the diffusion of Pb and I into V$_{MA}$ (Supplementary Fig. 14) while the other is caused by the PbI$_6$ octahedron slipping from corner sharing to edge sharing. Finally, the structure gradually evolves to PbI$_2$. **Fig. 4 e-g** illustrate the atomic-scale structural evolution and the related ion migration during the decomposition from MAPbI$_3$, MA$_{0.5}$PbI$_3$, to the final PbI$_2$, mainly involving two processes, forming V$_{MA}$ and the collapse of perovskite structure via diffusion of Pb-I.

**Discussion and summary**

The extreme beam sensitivity of OIHPs hinders atomic-resolution imaging and thus the detailed investigations on their structure-property relationships. By DDEC camera, we have determined that the threshold doses for superstructure formation is about 2.7 e Å$^{-2}$, and perovskite collapses into PbI$_2$ within 272.0 e Å$^{-2}$, both of which are smaller than those measured by electron diffraction (ED)[19]. This is mainly because ED pattern is obtained from a comparably larger region of 300-800 nm and represents the average information. These threshold conditions can guide future TEM characterizations and encourage more atomic-scale investigations about OIHPs.

Atomic-scale imaging of MAPbI$_3$ and its degradation pathway allows us to better understand properties of OIHPs. For example, the observed off-center displacements (up to ~ 30 pm) between different atom columns likely indicate the polar nature of this material[36], although further studies are needed to fully clarify this point, including determining MA configuration[15] and quantifying the effects from possible mistilt and residual aberrations[37]. The superstructure phase with additional reflections has been previously reported to be likely related with octahedra tilts[18] or ordered iodine vacancies[19] based on reciprocal-space ED analysis, our atomic-resolution imaging, however, has suggested that cation-ordered vacancies are more likely. Furthermore, the proposed two-step degradation pathway is initiated with the loss of MA to form MA$_{0.5}$PbI$_3$ and followed by the diffusion of Pb and I with perovskite structure collapsing into 6H-PbI$_2$. Interestingly, such an intermediate phase (MA$_{0.5}$PbI$_3$) with locally ordered vacancies can stably exist before perovskite collapses, suggesting the degraded structure with partial MA loss may be recovered. This likely sheds light into reversible photoinduced structural changes without forming PbI$_2$[38]. Such self-healing behavior under illumination has also been observed in MAPbI$_3$-based solar cells[39]. In addition, loss of MA causes the increased bandgap, which provides a potentially new strategy to tune the band gap in constructing tandem solar cells[40]. Also, the increased bandgap facilitates multiwave electroluminescence emission and adjusting various color luminescence under increasing bias voltage[41,42].

What happens to organic compositions during the degradation is still under debate. Some suggest that N-H bond is broken under illumination[43], heat[44] and moisture[45], thus forming CH$_3$NH$_2$ and HI while others insist forming NH$_3$ and CH$_3$I with C-N bond destroyed under heat[46]. Our vibrational spectroscopy has revealed that both C-N and N-H bonds can be destroyed under irradiation, releasing NH$_3$ and leaving hydrocarbons, which provide new insights into understandings of degradation mechanism for MAPbI$_3$.

Ion migration in OIHP-based electronic device is regarded as one of the most important processes, which contributes to the phase segregation, hysteresis in J-V curves and device degradation[47]. Previous studies about ion migration are either based on calculations or macro-measurements[47] like time of flight secondary ion mass spectroscopy[48], conductive atomic force microscopy[49], and energy-dispersive X-ray mappings[50,51], all without achieving the atomic-scale resolution in real space. Our atomic-resolution imaging provides direct evidence for ion diffusion of MA, I, and Pb under electron beam irradiation, thus providing some insights into understanding ion-migration-induced phase transformations and degradation, and consequently optimization of device performance. For example, since the gentle irradiation under illumination likely only causes the reversible loss of MA with perovskite structure preservation, accordingly the device efficient can be fully recovered at early degradation stages[52]. However, longer irradiation brings in I and Pb diffusion to induce irreversible transformation into PbI$_2$, thus bringing in an irreversible device performance degradation. The irreversible performance decline has also been observed under elevated temperature[53,54] and large bias[55] due to irreversible ion migration and structure degradation.

In summary, we have acquired the atomic structure of MAPbI$_3$, determined the threshold doses during TEM characterizations, and clarified the atomic-scale ion migration during its degradation into PbI$_2$. The degradation pathway is proposed to be a two-step, initialed by loss of MA and followed by diffusion of Pb and I to form PbI$_2$, during which both C-N and N-H bonds can be destroyed under irradiation, releasing NH$_3$ and leaving hydrocarbons. Such degradation process leads to the gradual increase of bandgap. These findings can be used to guide the future TEM characterizations, enrich the understandings of the degradation mechanism and optimization strategies, and provide atomic-scale insights into understanding its fundamental properties.

**Methods**

**MAPbI$_3$ Synthesis.** MAPbI$_3$ nano-crystals were bought from Xiamen Luman Technology Co., Ltd. Micro MAPbI$_3$ was synthesized as previously reported[56]. PbI$_2$ and MAI were prepared in γ-butyrolactone (GBL) with molar ratio 1:1 and the concentration of 1.3 mol L$^{-1}$. Then they were stirred at 70 °C for 12 hours to obtain the precursor solution. After precursor solution was filtered using polytetrafluoroethylene (PTFE) filter with 0.22 μm pore size. Two pieces of fluorine-doped tin oxide (FTO)/TiO$_2$ substrates were clamped together and vertically and partially soaked in MAPbI$_3$ precursor solution (10 ml) at 120 °C. Then the precursor solution was added twice one day in the nitrogen glove box. After several days, the substrates with single-crystal MAPbI$_3$ film were brought out, and dried at 120 °C for 10 minutes in nitrogen.

**Characterization and image analysis.** ED patterns and HRTEM images were acquired at an aberration corrected FEI (Titan Cubed Themis G2) operated at 300 kV. The Cs value is ~ 6.8 μm. Before acquiring images, the illumination range was set to be 3 μm in diameter. To shorten the exposure time to electron beam, we adjusted the defocus and the astigmatism well in one 3-μm region, then blanked beam, and moved to another 3-μm region to acquire the HRTEM image. HRTEM images were acquired at a magnification of 77 k by DDEC camera in electron-counting mode with the dose fractionation function. The correction of drift is achieved by using the DigitalMicrograph software by cross-correlation. The original image contains 40 subframes in 4 s and every 2 subframes were summed up to enhance the contrast for a more accurate estimation of drift. Hanning window and Bandpass filters were combined to improve the accuracy of the cross correlation. HRTEM images in Fig. 1, Fig. 2e, Fig. 4, and Supplementary Fig. 2, 6, 7 have been ABSF-filtered. HRTEM images in Fig. 2a, c have been first ABSF-filtered and then averaged from multiple regions using a home-made MATLAB code to reduce noise.

The morphology was characterized by SEM (FEI Quanta 200F) and CL spectrum was acquired using Rainbow-CL of Beijing Goldenscope Technology Co., Ltd at 5 kV, spotsize 4. Each single CL spectrum was acquired using 4 s. Vibrational spectroscopy was obtained under 30 kV at Nion U-HERMES200. Each spectrum was stacked from 200 single spectra, obtained using 800 ms, and the processing details were shown in Supplementary Fig. 12. The simulated ED patterns were obtained by the SingleCrystal (Crystalmaker) software. Structural models were acquired using Vesta software.

**Density functional theory calculation**. Our first-principles calculations were performed within the framework of density functional theory as implemented in the Vienna ab initio simulation package[57,58]. The ion-electron interaction was depicted by projector augmented-wave method[57,58]. The electron exchange correlation was treated by the generalized gradient approximation with Perdew-Bruke-Ernzerhof functional[60]. A kinetic cutoff energy was set as 500 eV for the Kohn-Sham orbitals being expanded in the plane-wave basis. The atomic positions were fully optimized with a conjugate gradient algorithm until the Hellman-Feynman force on each atom are less than 0.01 eV/Å[60]. The Monkhorst-Pack k- point meshes was sampled as 9×9×7[62].

**Ab initio molecular dynamics simulation**. We performed the ab initio molecular dynamic (AIMD) simulation. The plane-wave cutoff was set as 500 eV and the Brillouin zone is sampled at the Γ point. The AIMD was performed in the canonical ensemble at 300 K.

**Data availability.** The authors declare that all relevant data are included in the paper and its Supplementary Information files. Additional data including the codes are available from the corresponding author upon reasonable request.

**Acknowledgements**

This work was supported by the National Key R&D Program of China (2019YFA0708200), the National Natural Science Foundation of China (11974023, 52021006, 11772207), the Key R&D Program of Guangdong Province (2018B030327001, 2018B010109009), and the "2011 Program" from the Peking-Tsinghua-IOP Collaborative Innovation Center of Quantum Matter, the Guangdong Provincial Key Laboratory Program from the Department of Science and Technology of Guangdong Province (2021B1212040001), Natural Science Foundation of Hebei Province for distinguished young scholar (A2019210204) and Hebei Provincial Department of science and technology central guiding local science and technology development fund projects (216Z4302G). The authors gratefully acknowledge the Electron Microscopy Laboratory at Peking University for the use of electron microscopes and the support of the Center for Computational Science and Engineering at Southern University of Science and Technology. The authors also thank Mr. Yunkun Wang and Prof. Yunan Gao at Peking University for Raman test of MAPbI$_3$ and Prof. Yuan Yao at Institute of Physics, Chinese Academy of Sciences and Dr. Wenquan Ming at Hunan University for the good suggestions about the HRTEM data processing.

**Author contributions**

P. Gao, J.Y. Li, and J.J. Zhao conceived and supervised the project. S.L Chen performed TEM experiments and analyzed experimental data with the direction of P. Gao. and help from J.L. Qi and J. Cao. C.W. Wu performed the calculations under the guidance of X. Wang. B. Han carried out vibrational spectroscopy and Z.T. Liu performed CL measurements. Z. Mi and W.Z. Hao grew MAPbI$_3$ crystals under the guidance of J.J. Zhao. Q. Zhang, J.C. Feng, K.H. Liu and D.P. Yu provided additional specimen. S.L. Chen, J.Y. Li and P. Gao wrote the manuscript and all authors participated in the revision.

**Competing interests**

The authors declare no competing interests.

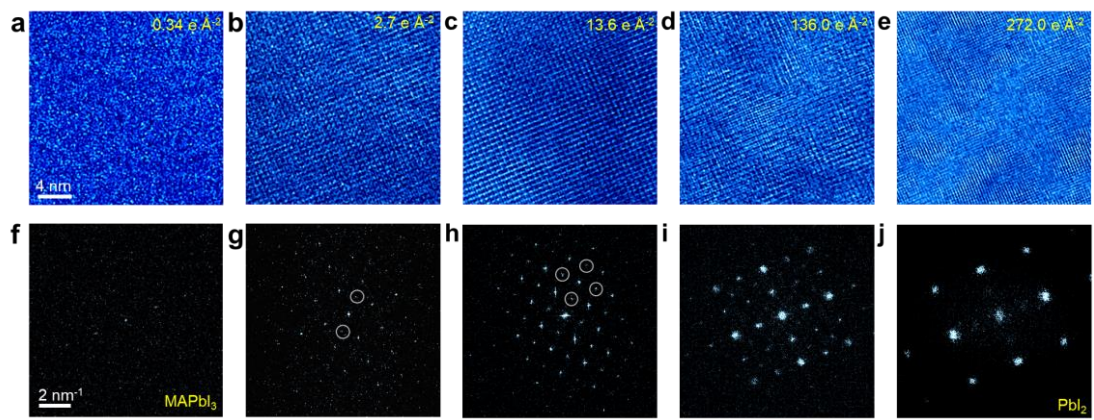

**Fig. 1 Tracking structure evolution during the decomposition of MAPbI3. a-e** Time-series HRTEM images during the degradation of MAPbI3 under electron beam irradiation. The corresponding doses are marked on each panel. **f-j** The corresponding FFT patterns from MAPbI3 to 6H-PbI2. Circles indicate the superstructure reflections.

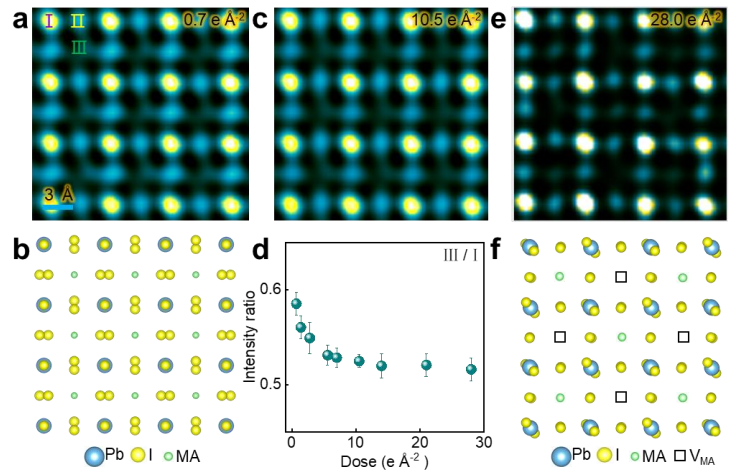

**Fig. 2 Atomic-imaging of loss of MA and intermediate phase. a** HRTEM image acquired at 0.7 e Å$^{-2}$. 'I', 'II' and 'III' columns correspond to Pb-I, I, and MA atomic columns respectively. **b** Structural model of tetragonal MAPbI$_3$. **c** HRTEM image acquired at 10.5 e Å$^{-2}$. **d** Intensity ratio of 'III' to 'I' atomic column with increased dose. **e** HRTEM image acquired at 28.0 e Å$^{-2}$. **f** Structural model of relaxed MA$_{0.5}$PbI$_3$. The squares indicate ordered MA vacancies.

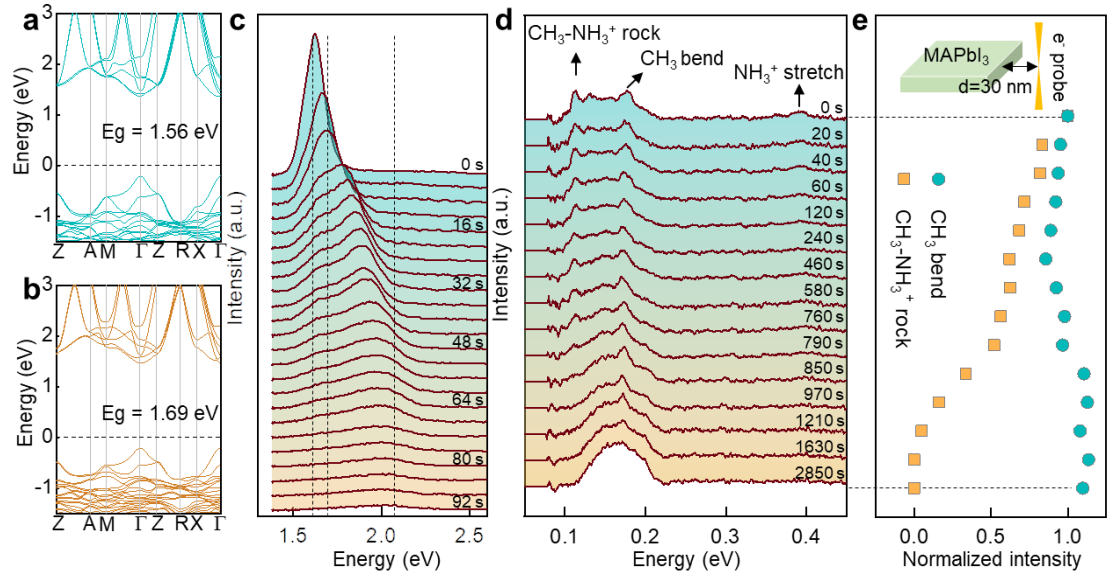

**Fig. 3 Electronic structure and chemical bonding evolutions during the degradation. a, b** Electronic structures of MAPbI$_3$ and MA$_{0.5}$PbI$_3$. The Fermi level is set to zero. **c** Time-series CL spectra showing the bandgap gradually increases from 1.6 eV to 2.05 eV. The dashed lines show the calculated bandgaps at 1.56, 1.69 and 2.15 eV for MAPbI$_3$, MA$_{0.5}$PbI$_3$ and 6H-PbI$_2$, respectively. **d** Time-series vibrational spectroscopy under 'aloof' mode with the electron probe 30 nm away from MAPbI$_3$. Black arrows indicate the peaks related with CH$_3$-NH$_3^+$ rock, CH$_3$ bend and NH$_3^+$ stretch. The background was subtracted by power-law function. **e** Evolutions of normalized intensities of CH$_3$ bending and CH$_3$-NH$_3^+$ rock during the degradation. Inset is the schematic diagram of 'aloof' mode showing the electron probe is about 30 nm away from MAPbI$_3$ for vibrational spectroscopy measurement.

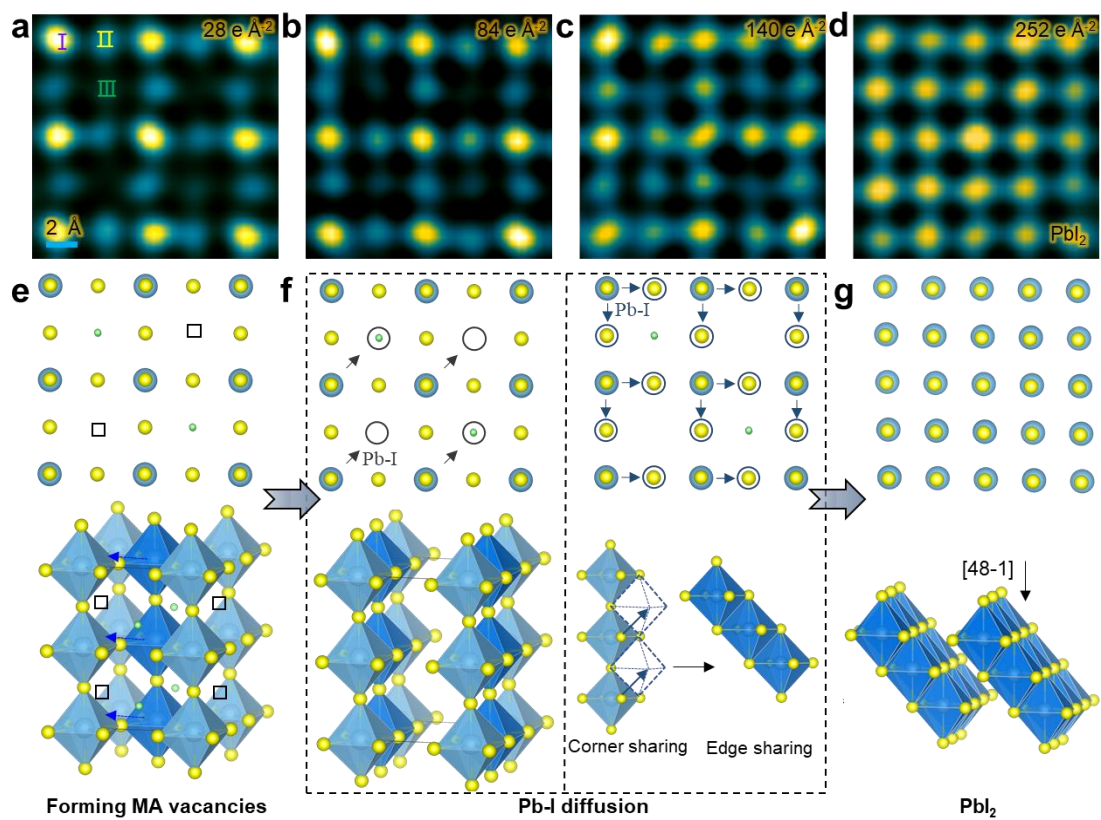

**Fig. 4 Atomic-scale imaging the decomposition pathway. a-d** HRTEM images with increased doses during the degradation into PbI$_2$. The corresponding doses are marked on each panel. **e** Atomic structure of MA$_{0.5}$PbI$_3$. Black squares indicate V$_{MA}$. **f** Atomic structures to illustrate two kinds of Pb-I diffusion. Left panel shows Pb-I diffuses to MA while the right panel illustrates the PbI$_6$ octahedron slipping from corner sharing to edge sharing. **g** Atomic structure of final product PbI$_2$.